\documentclass[a4paper,twoside]{article}
\usepackage{graphicx, subcaption, calc, amssymb, amstext, amsmath, amsthm, multicol, pslatex, apalike, algorithm2e, footmisc, SCITEPRESS, hyperref}
\usepackage[T1]{fontenc}
\usepackage[utf8]{inputenc}
\usepackage{float, comment, afterpage, multirow, longtable, adjustbox, tabularx, changepage, url, ulem, array, booktabs, stackrel, pgfplots, pgf-pie, balance, fontawesome5}

\usepackage{tikz}
\usepackage[most]{tcolorbox} 

\pgfplotsset{compat=1.7}

\newtcolorbox{takeawaybox}[1]{
    colback=white,
    colframe=gray!30!white, 
    colbacktitle=gray!10!white, 
    coltitle=black,
    fonttitle=\bfseries,
    fontupper=\small,
    title={\large\faBook~#1},
    boxrule=1pt,
    sharp corners,
    enhanced,
    attach boxed title to top left={
        yshift=-\tcboxedtitleheight/2,
        xshift=1mm,
        yshifttext=-\tcboxedtitleheight/2,
    },
    top=0mm,
    bottom=0mm,
    left=3mm,
}

\newcommand{\blackcircle}[1]{%
    \begin{tikzpicture}[baseline=-0.75ex]
        \node[circle, fill=black, inner sep=1.5pt, text=white, text width=0.5em, align=center] {#1};
    \end{tikzpicture}%
}

\begin{document}
\title{ChatGPT as a Software Development Bot: A Project-based Study}
\author{
  \authorname{Muhammad Waseem\sup{1}, Teerath Das\sup{1}, Aakash Ahmad\sup{2},\\ Peng Liang\sup{3}, Mahdi Fahmideh\sup{4}, Tommi Mikkonen\sup{1}}
 \affiliation{
    \sup{1}Faculty of Information Technology, University of Jyväskylä, Jyväskylä, Finland \\
   \sup{2}School of Computing and Communications, Lancaster University Leipzig, Leipzig, Germany \\
    \sup{3}School of Computer Science, Wuhan University, Wuhan, China \\
    \sup{4}School of Business at University of Southern Queensland, Queensland, Australia
  }
  \email{\{muhammad.m.waseem, teerath.t.das, tommi.j.mikkonen\}@jyu.fi, a.ahmad13@lancaster.ac.uk, liangp@whu.edu.cn, mahdi.fahmideh@unisq.edu.au}
}


\abstract{Artificial Intelligence has demonstrated its significance in software engineering through notable improvements in productivity, accuracy, collaboration, and learning outcomes. This study examines the impact of generative AI tools, specifically ChatGPT, on the software development experiences of undergraduate students. Over a three-month project with seven students, ChatGPT was used as a support tool. The research focused on assessing ChatGPT's effectiveness, benefits, limitations, and its influence on learning. Results showed that ChatGPT significantly addresses skill gaps in software development education, enhancing efficiency, accuracy, and collaboration. It also improved participants' fundamental understanding and soft skills. The study highlights the importance of incorporating AI tools like ChatGPT in education to bridge skill gaps and increase productivity, but stresses the need for a balanced approach to technology use. Future research should focus on optimizing ChatGPT's application in various development contexts to maximize learning and address specific challenges.}

\maketitle

\section{Introduction}
\label{sec:Introduction}
 
Artificial Intelligence (AI) and especially Machine Learning (ML) is increasingly being integrated into software development processes \cite{barenkamp2020applications}, leveraging ML algorithms and intelligent systems to automate code generation, bug fixing, and testing, thereby enhancing efficiency and reducing manual effort, \cite{shehab2020aiam}. This technological progression paves the way towards Large Language Models (LLMs) \cite{kim2021scalable},  which have revolutionized textual and coding tasks by understanding and generating human-like text. Notable developments within LLMs include BERT \cite{devlin-etal-2019-bert}, GPT, and LLaMA, each having distinct functionalities. Among these, the Generative Pre-trained Transformer model, for example  GPT-3 has garnered substantial attention due to its ability to train on extensive human data and generate coherent text, both labelled and unlabelled datasets.

\textbf{Context and Motivation}: GPTs have pervaded numerous applications, notably AI chatbots, content creation, and diverse areas within software development, as demonstrated by multiple studies and practical applications (e.g., \cite{ahmad2023towards}). Nevertheless, empirical reports specifically detailing the application of GPTs, including GPT-3 and the more recent ChatGPT, in the training of undergraduate students across various software development phases remain notably limited. We will use 'students' for short in the rest of the paper.  For instance, GPT-3 has been investigated for its capabilities in automatic code generation and automated code documentation, and utilized to develop tools like \textit{Codex}, which converts natural language instructions into code and aids in automatically generating code documentation \cite{khan2022automatic}. Simultaneously, students consistently face challenges such as understanding fundamental architectural concepts, managing subsystem integration, and confronting debugging issues. A survey involving both educators and students highlighted common struggles that students face, including program designing, procedure segmentation, and error identification within their code, while additional research has shown the occurrence of recurrent code quality issues in novice-authored programs \cite{8103449} and difficulties in utilizing tools like version control systems, performing infrastructure testing, and managing bug tracking and resolution.
\textbf{Objective of the Study}: The objective of this study is to explore the effectiveness of ChatGPT as software development bot within different phases of the software development life cycle, as undertaken by students. This includes evaluating ChatGPT's impact on requirements analysis, design, architecture, development, testing, and deployment. The study also aims to identify ChatGPT's advantages and limitations in each phase, understand its influence on students' learning curves and skill development, assess improvements in software development proficiency, and pinpoint challenges faced by students. By considering objective we formulated following four Research Questions (RQs):
\begin{itemize}
    \item \textbf{RQ1}: How did the utilization of ChatGPT as development bot impact the requirements analysis, design, architecture, development, testing, and deployment phases of software development, and what are the advantages and limitations of its use at each phase of the software development life cycle?
    \item \textbf{RQ2}: How did the involvement in software development projects that incorporated ChatGPT influence the learning curve and skill development of the students?
    \item \textbf{RQ3}: To what extent did the students' proficiency in software development concepts improve through their participation in these projects with the help of ChatGPT as development bot?
    \item \textbf{RQ4}: What challenges did the students face when using ChatGPT as development bot?
\end{itemize}

The \textbf{contributions} of this study are (i) assessing ChatGPT's impact on students' software development lifecycle, with a focus on advantages and limitations; (ii) identifying challenges and evaluating ChatGPT's educational value in student software projects; and (iii) presenting evidence of ChatGPT's integration and effectiveness in software development, contributing to discussions on human-AI collaboration. 

\section{Related Work}
\label{sec:RelatedWork}
\subsection{ChatGPT-Assisted Software Design} 
Bencheikh et al.\cite{bencheikh2023exploring} conducted a study that explores the efficacy of AI tools, particularly ChatGPT, in generating software requirements efficiently. They highlighted the ability of ChatGPT to emulate human expertise, though emphasizing the indispensable nature of human feedback to enhance requirement quality. Moreover, the time efficiency of ChatGPT was acknowledged, albeit with a recognition that experienced human participants tend to produce more comprehensive requirements. The study also differentiated between the premium and free versions of ChatGPT, showing a superior consistency and overall quality in the former. On a related note, White et al.\cite{white2023chatgpt} proposed several ChatGPT prompt patterns to elevate software requirements elicitation. Transitioning to the architectural phase of software development, a case study by Ahmad et al. \cite{ahmad2023towards} shed light on the collaborative efforts between a novice software architect and ChatGPT.

\subsection{Code Generation and Testing with ChatGPT} 
In the coding and implementation, several researchers have used ChatGPT. For instance, Al-Khiami et al.\cite{al2023leveraging} conducted a case study to explore the feasibility of using ChatGPT for generating JavaScript code suitable for Android Studio, aimed at creating a functional app. Concurrently, Bera et al.\cite{bera2023use} employed ChatGPT to support agile software development. Additionally, the significance of ChatGPT prompt patterns in improving code quality and facilitating refactoring was discussed by White et al. \cite{white2023chatgpt}. Beyond these, the GPT-3 model has been utilized for automatic code generation \cite{narasimhan2021cgems} and automation of code documentation \cite{khan2022automatic}, with Tian et al.\cite{tian2023chatgpt} employing ChatGPT as a programming assistant, albeit noting limitations regarding ChatGPT's attention span. 
\section{Research Method}
\label{sec:ResearchMethod}

\begin{figure}[t]
    \centering
    \includegraphics[width=2\columnwidth, height=0.25\textheight, keepaspectratio]{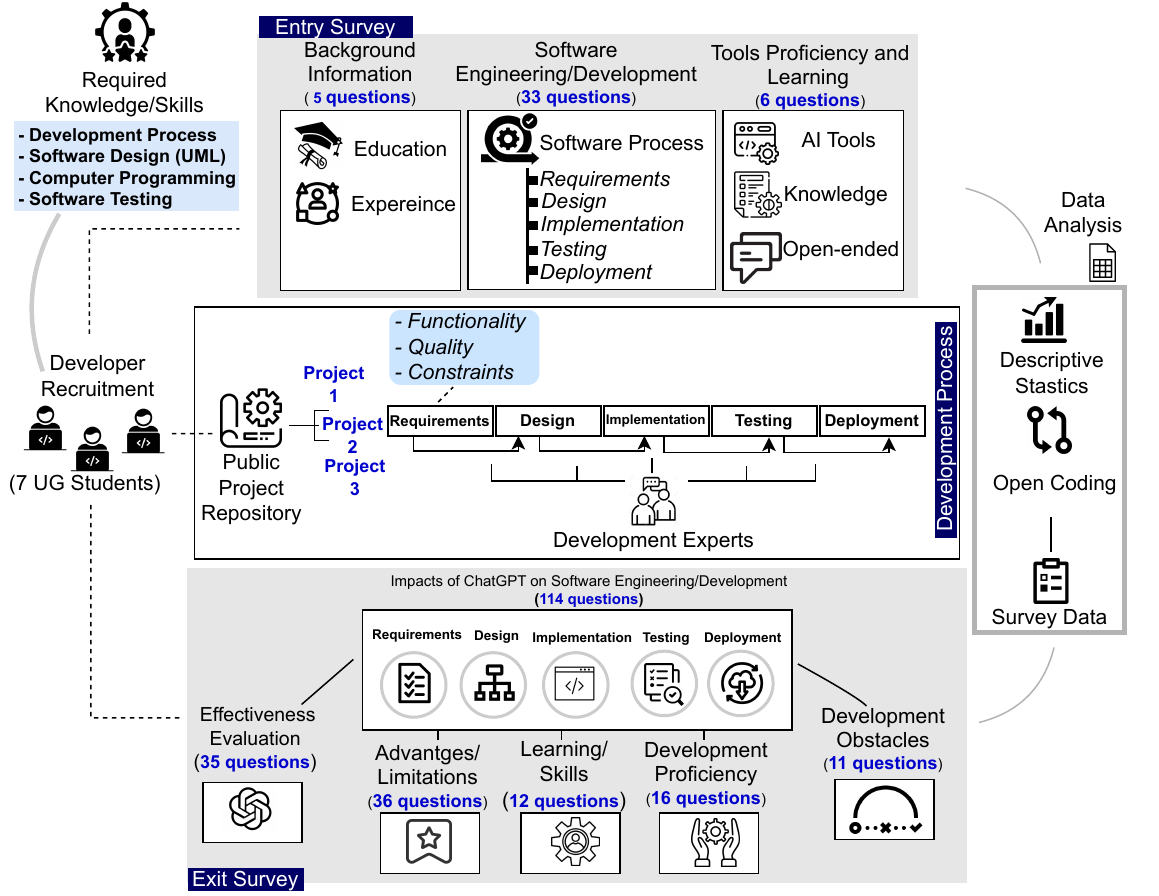}
    \caption{Overview of the Research Method}
    \label{fig:methodology}
\end{figure}

The research method of this study consists of four phases, illustrated in Figure \ref{fig:methodology} and detailed as follows. 

\subsection{Developers and Projects Selection}
\label{DeveloperProjectsSelection}
 \textbf{Developers Selection}: Seven undergraduates from IT, computer science, and software engineering programs were selected to evaluate their software development experiences with ChatGPT (versions GPT-3.5 and GPT-4), recruited through university's online bulletin board and social media, followed by interviews. These first or second-year students were chosen for their limited familiarity with software development practices, including Scrum, CI/CD, UML, Python programming, automated testing, and deployment.

\textbf{Project Selection}: This study focus on three publicly announced software projects assigned to a group of seven students, divided into two teams. ChatGPT, an AI-powered tool, was utilized extensively to support students throughout the software development process, spanning from understanding project requirements to deployment. 

We developed three systems to enhance procurement processes, foster educational technology collaboration, and assess student performance. (i) For Solita Ltd, the AI Procurement Assistant (AIPA) utilized AI to streamline identifying procurement opportunities, with ChatGPT simplifying procurement option identification. (ii) The AI-based Teacher-Tech Forum, created for Jyväskylä University of Applied Sciences, established an online platform for teachers and tech companies to exchange ideas for using technology in education. (iii) Lastly, the AI-based Skills Assessment SaaS, aimed at city governments through the AXZ public procurement system, assessed junior-level students' performance in subjects like mathematics and English, providing municipalities and schools with valuable insights into student achievements.

\subsection{Conducting Entry Survey}
In this study, we adopted a cross-sectional survey design, recognized for its effectiveness in capturing data at a specific point in time across a targeted population \cite{kitchenham2008personal}. Opting for a web-based approach facilitated an efficient, cost-effective method of data collection in diverse formats \cite{lethbridge2005studying}. The survey's design was informed by both grey and peer-reviewed literature, alongside consultations with software engineering professors possessing industrial experience. Our team, leveraging academic and industrial expertise, refined the survey through focus group discussions, ensuring its alignment with the study's objectives. Notably, a pilot survey was conducted with two students external to the development team to validate the survey instrument's relevance and effectiveness. The survey instrument, detailed in the Entry Survey Questionnaire sheet \cite{Dataset}, comprised three sections with a total of 53 questions, designed to explore the participants' backgrounds, software engineering experiences, and proficiency with AI tools, including their readiness to learn these tools.

\subsection{Development Process}
The development process began with identifying projects from Finland's public procurement system. One project was chosen from Solita Oy software development company, one from Jyväskylä University of Applied Sciences (JAMK), and one from the procurement system, following discussions between the study authors and developers. The students were distributed into two teams, working on their projects in parallel. Team roles, like front-end and back-end developers, were predefined. After selecting the projects, the developers engaged in iterative activities, such as collecting requirements, setting up CI/CD infrastructure, software design, project development, system testing, and system deployment, in collaboration with client representatives from Solita Oy and JAMK University.

\subsection{Conducting Exit Survey}
The exit survey, similar to the entry survey, used a cross-sectional approach \cite{kitchenham2008personal} (detailed in the Exit Survey Questionnaire sheet \cite{Dataset}) with 114 questions. It aimed to thoroughly understand ChatGPT's impact on software development, assessing its effectiveness, advantages, limitations, educational value, challenges, and its potential to improve project outcomes and developer skills. Tailored to address the study's Research Questions, it used a 5-level Likert scale for responses. The survey is organized into five sections, each focusing on different aspects of ChatGPT's impact in software development projects.

\subsection{Analysis}
We employed descriptive statistics to analyze the quantitative (i.e., closed-ended questions). To evaluate the responses to few open-ended questions, we employed open coding from grounded theory \cite{glaser2017discovery} to segment and label the survey data, thereby identifying and emphasizing the text fragments in each student's response that could be treated as distinct data points regarding ChatGPT.
\section{Results}
\label{Sec:Results}

\subsection{Background, Experience, and Readiness}

We present below the key results from the survey.

\begin{figure}
    \centering
    \scriptsize
    \includegraphics[width=5.1\textwidth, height=0.62\textheight, keepaspectratio]{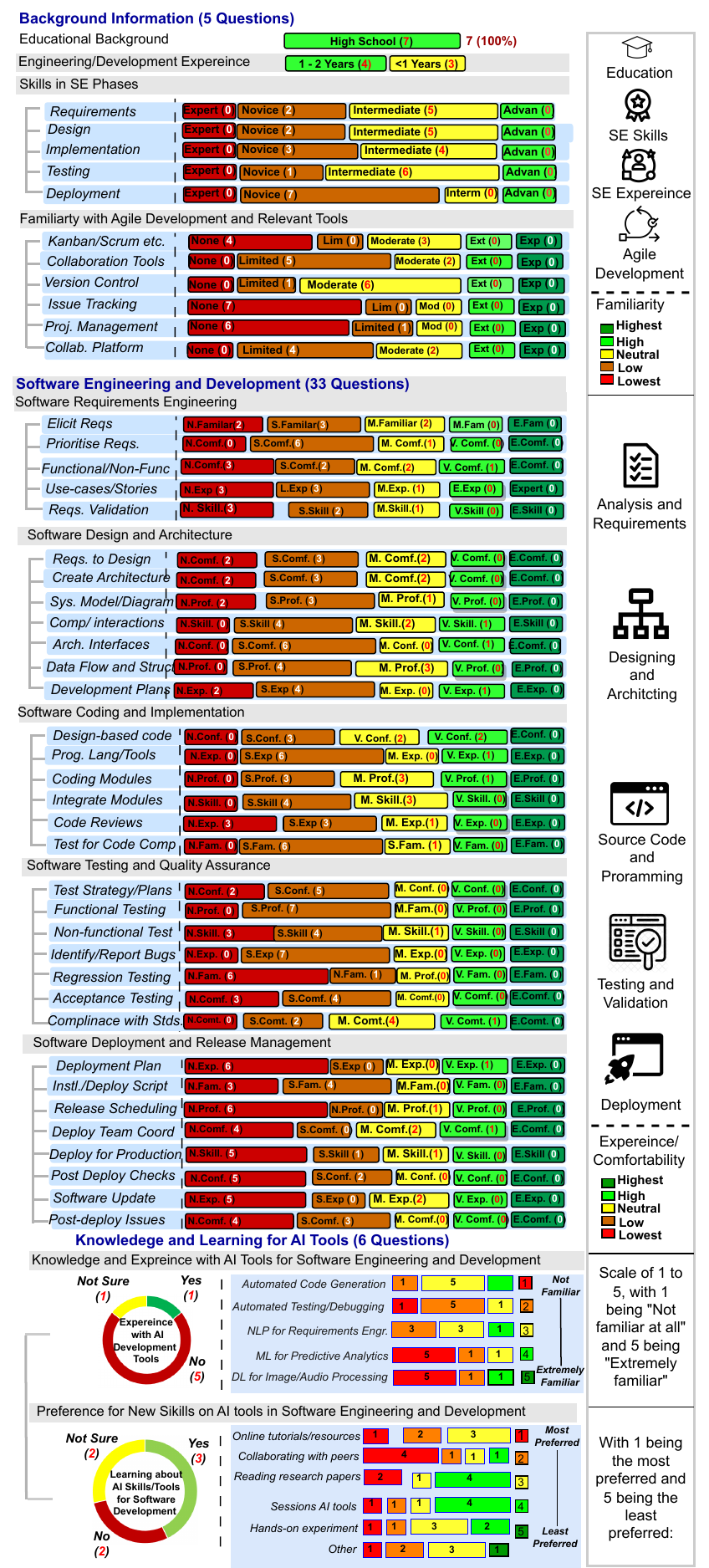} 
   \caption{Results Overview: Background, Experience, Readiness}
    \label{fig:EntrySurvey}
\end{figure}

\textbf{Developers' Background}: The survey data revealed that the participants mostly had high school education and limited software development experience (see Figure \ref{fig:EntrySurvey}): four with 1-2 years and three with less than a year, related to undergraduate computing studies. Most self-assessed their software engineering skills as `Intermediate', indicating a lack of advanced expertise. There was also a noticeable unfamiliarity with agile methodologies and limited proficiency in professional collaboration tools like Jira, suggesting the participants lacked the typical educational and practical skills for advanced software development roles.

\textbf{Experience in Software Engineering}: The entry survey with 35 questions on the software development life cycle showed a consistent trend (Figure \ref{fig:EntrySurvey}). Participants were mostly `Somewhat familiar' with Software Requirements Engineering tasks and `Somewhat comfortable' in Development and Design phases. Confidence in Implementation, Testing, and Quality Assurance ranged from `Somewhat' to `Moderately confident`, with varying proficiency in coding and test strategy development. The largest skill gap was in Deployment and Release Management, where most had `No experience' or were `Not proficient at all', underscoring significant skill gaps, particularly in Deployment and Release Management.

\textbf{Readiness to Learn about AI tools}: Figure \ref{fig:EntrySurvey} outlines participants' AI tool proficiency and learning readiness in software development. Of seven respondents, five had minimal AI tool usage, with some familiarity through ChatGPT. They showed higher proficiency in Machine Learning for Predictive Analytics and Deep Learning for Image/Audio Processing. The preferred learning methods for AI skills were hands-on experimentation and projects, followed by online tutorials or video courses. Collaboration on AI projects was also favored, indicating a preference for collaborative learning. Less preferred were reading research papers and attending workshops or seminars on AI tools.

\textbf{Perception and Potential of AI Tools}: Open-ended questions explored students' views on AI tools in software development, covering aspects like enhancement, limitations, and future prospects. We summarize participants' responses (P1 to P7) as follows:

(i) Students perceived AI tools like ChatGPT as beneficial, especially for complex projects with limited experience. They appreciated AI's quick code generation and efficiency in finding tools and algorithms, reducing manual search time (\faUser P3, \faUser P4, \faUser P5). However, concerns about the quality of rapid outputs and integration difficulties were noted (\faUser P1, \faUser P2). While AI offers productivity and efficiency, quality assurance and system integration remain challenges. 

(ii) Concerns included AI's impact on privacy, ethics, developer education, and reliability. Risks of privacy breaches and sensitive information leaks in AI-generated code were mentioned (\faUser P1, \faUser P2), as well as ethical concerns over code ownership (\faUser P4). Over-reliance on AI possibly hindering skill development (\faUser P3) and AI code's reliability (\faUser P7) were other concerns.

(iii) Mixed feedback was given on AI's integration in software development. Concerns about AI's long-term sustainability (\faUser P1), the development of AI tools for larger project assistance (\faUser P2), trust issues due to past inaccuracies (\faUser P5), and AI's potential in streamlining workflows for new developers (\faUser P6) were discussed. Feedback ranged from acknowledging AI's benefits to addressing challenges in trust and effective utilization.

\begin{takeawaybox}{Takeaways}
\scriptsize
\textbf{Significant Skill Gaps}: The surveyed developers showed notable gaps in their software engineering skills, especially in deployment and release management stages.

\textbf{Eagerness to Learn AI}: Despite minimal previous experience with AI tools, participants demonstrated a strong desire to learn, particularly through hands-on experiences and projects.
\end{takeawaybox}

\begin{figure*}[t]
    \centering
    \includegraphics[width=6.25\columnwidth, height=0.61\textheight, keepaspectratio]{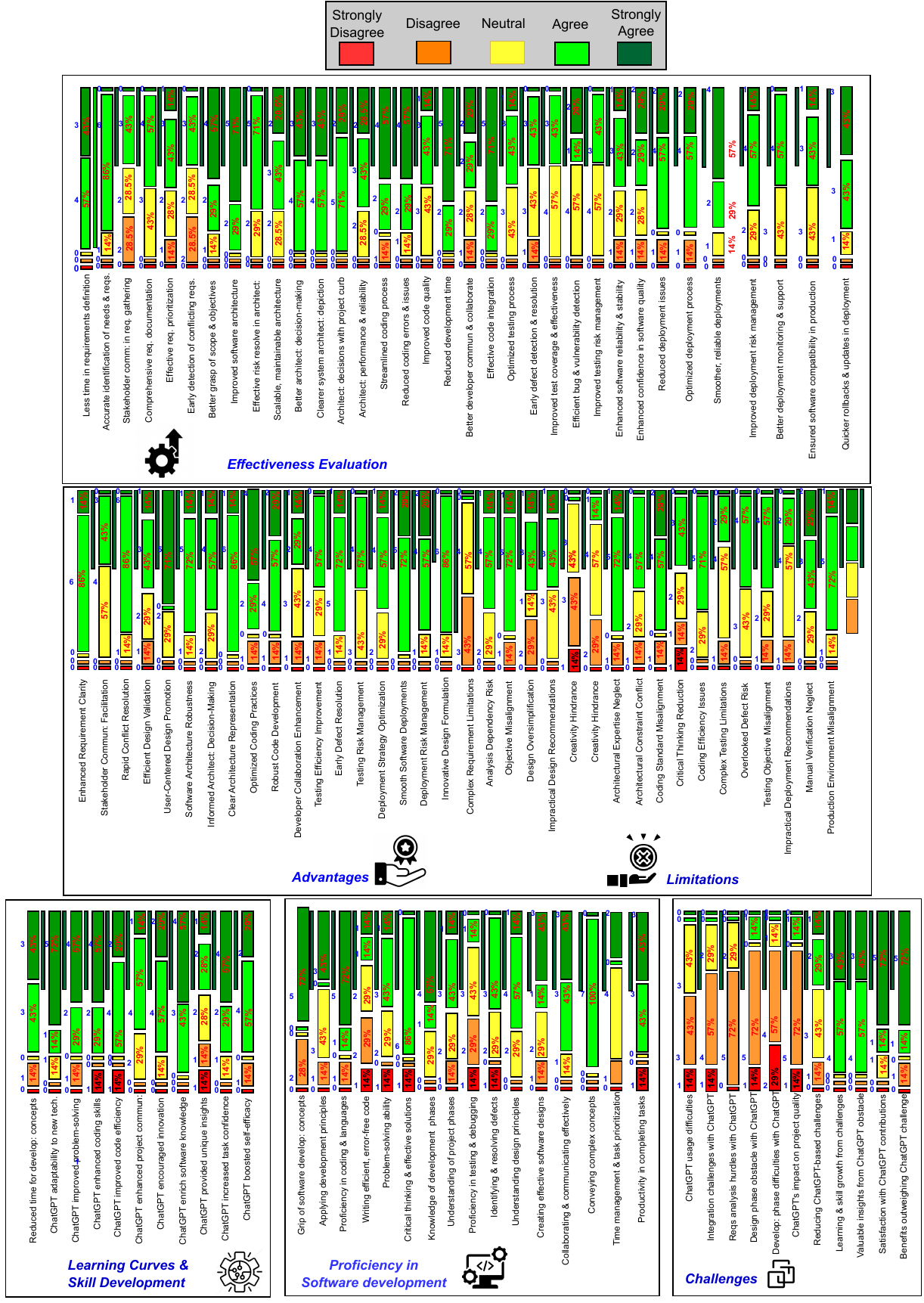}
    \caption{Results Overview: ChatGPT's effectiveness, Advantages, Limitations, Skill development, Proficiency}
    \label{fig:ExitSurvey}
\end{figure*}

\subsection{Effectiveness, Advantages, and Limitations (RQ1)}
To answer the RQ1, we asked 71 survey questions to gather answers from students who extensively used ChatGPT for three months to develop three systems as desribed in Section \ref{sec:ResearchMethod}.

\textbf{Effectiveness}: Developers reported (see Figure \ref{fig:ExitSurvey}) that ChatGPT's positive impact on software development's early stages. In requirements analysis, 7 respondents agreed ChatGPT significantly reduced time for defining requirements and 6 agreed it helped accurately identify needs. In design, 5 respondents agreed ChatGPT led to clearer, scalable architectures and 7 agreed it aligned architectural decisions with project constraints. ChatGPT also streamlined the development phase, improving coding process and quality as agreed by 6 students. In testing, it optimized processes, enabled early defect detection, and improved test coverage, with 3 to 6 respondents agreeing on its effectiveness in bug identification, risk management, and software reliability.

\begin{takeawaybox}{Takeaways}
\scriptsize
\textbf{Proficiency in Software Development}: ChatGPT consistently enhanced foundational understanding and soft skills in software development among participants, but experiences varied notably in its ability to provide unique insights during the software development process.
\end{takeawaybox}

\textbf{Advantages}: Developer responses (see Figure \ref{fig:ExitSurvey}) highlighted the perceived benefits of using ChatGPT in software development. Seven students agreed that ChatGPT enhanced requirement clarity. Its role in stakeholder communication was mixed, with 4 neutral and 3 in agreement. Opinions on its impact in conflict resolution, innovative design formulation, and efficient design validation were diverse. However, promoting user-centered design, robust software architecture, and informed architectural decision-making received positive feedback, with most students seeing advantages. Six students strongly agreed on the benefits of clear architecture representation. 

\begin{takeawaybox}{Takeaways}
\scriptsize
\textbf{Positive Impact}: ChatGPT was generally perceived to positively impact various facets of software development learning and skill development, despite some isolated instances of dissent or neutral perspectives from participants.
\end{takeawaybox}

\textbf{Limitations}: The survey addressed ChatGPT's limitations in software development. Opinions on complex requirement limitations were mixed, with 3 students disagreeing and 4 neutral. Analysis dependency risk saw varied responses (4 agreeing, 2 neutral). Objective misalignment in design was mostly neutral among 5 students. Design oversimplification and impractical recommendations received divided opinions, leaning slightly towards agreement. Creativity hindrance concerns were notable, with a significant portion disagreeing. Mixed responses were seen in architectural expertise neglect and constraint conflict. 

\subsection{Learning Curves and Skill Development (RQ2)}
To get the answer of RQ2, we asked 11 survey questions to gather students' opinions on the influence of ChatGPT on their learning and skill development. The survey results highlight the participants' experiences and perceptions regarding several aspects of using ChatGPT in their software development projects (see Figure \ref{fig:ExitSurvey}). For instance, when asked about ``Reducing the time needed to understand software development concepts'', the feedback was mostly positive: 3 participants agreed, and 3 strongly agreed. Similarly, ``Enhancing adaptability to new technology'' was positively received with 5 strongly agreeing and 1 agreeing.

\begin{takeawaybox}{Takeaways}
\scriptsize
\textbf{Valuable Learning Despite Challenges}:Participants unanimously agreed that they got valuable learning and insights from these experiences, suggesting that obstacles faced were constructively impactful on their developmental journey.
\end{takeawaybox}

\subsection{Proficiency in Software Development (RQ3)}

An agreed sentiment of approval was observed concerning ChatGPT’s role in enriching software knowledge (see Figure \ref{fig:ExitSurvey}), with all 7 respondents either agreeing or strongly agreeing. This indicates a consistent acknowledgment of ChatGPT’s capacity to enhance foundational understanding and proficiency in software development among the participants. Further, the efficacy of ChatGPT was evident in several other facets of the software development learning process. For example, 6 out of the 7 respondents concurred (agree or strongly agree) that the tool positively impacted their adaptability to new technologies, boosted practical problem-solving skills, and encouraged innovative thinking during their development.

On the other hand, it is noteworthy that certain aspects yielded more varied perceptions, providing a view of the participants' experiences. Specifically, regarding the statement, “ChatGPT provided unique insights,” there was an almost equal distribution of responses across all available options (strongly disagree to strongly agree), showcasing that the participants had diverse experiences and perceptions concerning ChatGPT’s capability in offering unique insights during the software development process. 

\subsection{Challenges (RQ4)}

In addressing RQ4, we asked 11 survey questions to understand the difficulties of students might have experienced while using ChatGPT in their projects. When it came to difficulties in various phases of project development with ChatGPT, the students generally did not find using ChatGPT hard or tricky. For example, in dealing with ChatGPT usage difficulties, 4 out of 7 respondents disagreed, suggesting that most found it user-friendly.

Even with the overall positive feedback regarding usability, there were areas where participants felt they learned and improved their skills from the challenges they faced. A key observation was that all participants (7 out of 7) agreed or strongly agreed that they experienced learning and skill growth from the challenges faced while using ChatGPT. In the same way, all respondents agreed or strongly agreed that they gained valuable insights from the challenges faced with ChatGPT. 
\section{Discussion}
\label{Sec:Discussion}

\blackcircle{1} \textbf{Positive Impact on Software Development Phases}: This study reported ChatGPT support for various software development phases, from requirements analysis to deployment, aligns with existing literature highlighting the beneficial role of AI in streamlining development workflows. For example, AI-driven tools have been substantiated for enhancing requirement gathering through natural language processing capabilities and facilitating design phases \cite{ahmad2023towards}. The positive impact across development phases prompts deeper explorations into specifying contexts, projects, or phases where ChatGPT’s application could be maximized.

\blackcircle{2} \textbf{Diverse Opinions on Advantages and Limitations}: The presence of both significant advantages and noticeable limitations in using ChatGPT for software development echoes previous literature (e.g., \cite{kalla2023study}) on AI tools in development environments. In terms of advantages, the recognized enhancement in requirement clarity and promotion of user-centered design correlate with the narrative of AI-driven tools being useful in translating user needs into technical requirements efficiently. The support in clear architecture representation and testing efficiency is coherent with studies that underscore automated testing and model-driven development (e.g., \cite{planas2021towards}) facilitated by AI. 
In contrast, the limitations, particularly around coding efficiency issues and design recommendation impracticalities, mirror concerns found in both in peered review and gray literature (e.g., \cite{codelogic2023}) that cautions against over-reliance on AI for complex decision-making in software development.
ChatGPT shows promise for enhancing parts of the Software Development Life Cycle (SDLC) but must be used wisely due to its limitations, like possibly oversimplifying complex tasks or suggesting unhelpful designs due to its partial understanding of context and software tasks. 

\blackcircle{3} \textbf{Positive Impact on Learning and Skill Development}: The majority of students found ChatGPT beneficial in aiding their understanding and skill enhancement in various facets of software development, which aligns with literaturesuggesting that AI can expedite the learning curve by providing instant, context-aware assistance  (e.g., \cite{al2023leveraging}). However, the dissenting responses, particularly the one participant who strongly disagreed about enhancing coding skills, indicating that personal experiences with ChatGPT varied. ChatGPT, acting as a supplementary tool, seems to align with these findings by providing support and instant feedback, enhancing both the theoretical and practical understanding of the students. However, it is crucial to consider the small sample size of 7 students, which may limit the generalizability of these findings. Positive impacts on skill development might be subjected to the initial proficiency level of the participants or other unaccounted contextual factors.

\blackcircle{4} \textbf{Varied Experiences about Software Knowledge and Soft Skills Development}: 
Even though ChatGPT positively impacted software knowledge and soft skills development, it provided varied experiences regarding unique insights into the software development process among participants. The varied experiences with ChatGPT mirror the existing literature on AI in education, which suggests that while AI can offer valuable support, the utility can differ based on users' expectations, existing skills, and the nature of tasks \cite{fraiwan2023review}. The disparity in how ChatGPT’s capability to provide unique insights was perceived implies that the tool might be interpreted or utilized differently among users. The limitation pertains to understanding the depth and nature of these varied experiences, given that the reasoning behind such divergence is not explored in detail. 

\blackcircle{5} \textbf{Valuable Learning Despite Challenges}: Challenges faced when using ChatGPT were not seen as drawbacks but rather as valuable learning experiences, aiding skill development and insight generation among students. This aligns with the pedagogical perspective that views challenges and obstacles as crucial learning elements that facilitate deep understanding and skills mastery \cite{ohlsson2014pedagogic}. Challenges faced, while initially perceived as hurdles, become opportunities for active learning and skill enhancement. Given the unanimously positive feedback regarding learning from challenges, there may be potential bias in the responses, or there might be a limitation in exploring the specific nature and impact of these challenges due to the small participant group and the quantitative approach of a survey. 

\section{Conclusions}
\label{Sec:Conclusion}
This study explored ChatGPT's impact on students in software projects as development bot, revealing skill gaps and a keen interest in AI. ChatGPT proved to be a beneficial support tool, enhancing efficiency and collaboration, but opinions varied on its advantages and limitations in design and coding. It improved participants' software development skills and soft skills, though its contribution to unique insights was inconsistent. Despite challenges, the experience was viewed positively for learning, underscoring AI tools' educational value.
Future research will explore into diverse experiences with ChatGPT as development bot, optimize its use with multi-agents, and create guidelines for effectively integrating AI in software engineering education at both undergraduate and graduate levels.

\bibliographystyle{apalike}
{\small
\bibliography{References}}

\end{document}